\documentclass[runningheads]{svmult}

\usepackage{makeidx}   
\usepackage{graphicx}  
\usepackage{subeqnar}  
\usepackage{multicol}  
\usepackage{physprbb}  
\makeindex             

\begin{document}
\title*{Blue Horizontal-Branch Stars\protect\newline 
and Simple Stellar Populations}
\titlerunning{Horizontal-Branch Stars \& SSPs}
%
\author{Hyun-chul Lee\inst{1,2}
\and Young-Wook Lee\inst{2}
\and Brad~K.~Gibson\inst{1}}
\authorrunning{Lee et~al.}
%
%
\institute{Centre for Astrophysics \& Supercomputing,
     Swinburne University,\\
     Hawthorn, Victoria, 3122, Australia\\
\and CSA, Yonsei University, Seoul 120-749, Korea}

\maketitle              

\begin{abstract}
The effects of blue horizontal-branch (HB) stars on the integrated 
spectrophotometric indices of simple stellar populations (SSPs) are addressed.
Synthetic spectra are drawn from a grid of simulated globular
clusters, constructed so as to reproduce the observed range of
HB morphologies in the Milky Way system.  Our models indicate that the
H$\beta$ line index increases by as much as 0.75~{\AA} 
and the integrated B-V colour becomes bluer by
upwards of 0.15~mag because of the contribution from blue HB stars.
We discuss the importance of both optical {\it and} far-UV colours as
age discriminants for SSPs, in particular for
high metallicities.  We suggest that relative age differences
(and consequently their differing HB morphologies) result in
non-linear colour-metallicity relations.
\end{abstract}

\section{Introduction}
There have been many efforts to develop evolutionary
population synthesis models (e.g. Bruzual \& Charlot 1993; Worthey 1994;
Buzzoni 1995; Vazdekis et~al. 1996; Maraston 1998; 
Lee et~al. 2000,2002) in order
to analyze the integrated spectrophotometric quantities of globular clusters
(GCs) and galaxies.  Ultimately, the primary use of such
software packages lies in attempting to derive the ages and metallicities
of stellar populations such that quantitative insights into the processses
governing galaxy formation can be uncovered.
Age and metallicity drive the integrated spectrophotometry of stellar
systems, and they also drive the resulting HB morphology of globular clusters.
It is interesting therefore to pose the question as to whether this underlying
morphology also manifests itself in the integrated spectrophotometry.
There is reason to suspect that
age is the {\it global} ``second parameter''
that controls HB morphology (after the ``first parameter'', metallicity -
e.g. Lee, Demarque \& Zinn 1994; Sarajedini, Chaboyer \& Demarque 1997;
Rey et~al. 2001; Salaris \& Weiss 2002), although for some GCs 
a third (or more) parameter may
be needed to explain their peculiar HB morphologies (such as the
blue tail phenomenon - e.g. Recio-Blanco et~al. 2002). 
In what follows, we describe the unique aspects of our recent models
which allow for an accurate determination of the role played by blue HB
stars in modifying\footnote{From that expected from the inclusion of
the main sequence turnoff and red giant branch phases {\it alone}.}
the spectrophotometric indices of SSPs.  The reader is directed to
Lee et~al. (2000,2002) for details.

\section{Wavy Features in the Optical}
We first construct a grid of simulated GCs spanning a range of HB morphologies 
- in order to match the observed morphology of inner halo Galactic GCs
(Galactocentric radius $\leq$~8~kpc) at their currently favoured mean age
($\sim$12~Gyr) and age dispersion ($\Delta${\it t}=0~Gyr) a Reimers (1975)
mass loss parameter $\eta$=0.65 was required.  The Salaris et~al.
(1993) $\alpha$-element enhancement correction was applied to our models.
The modified Gaussian mass distribution of Demarque et~al. (2000) -
with $\sigma$=0.02~M$_{\odot}$ - was used, shaping the widths of the
``wavy'' features seen in Figure~1.  The temperature-sensitive
B-V and H$\beta$ indices shown there illustrate the magnitude of the
contribution of blue HB stars (thick wavy lines - c.f.
Maraston \& Thomas 2000) within $\pm$4~Gyr of system's mean age.
The resulting wavy features stem primarily from our exploration of 
parameter space 2$-$4~Gyrs {\it older} than the Milky Way GCs.
The location of these wavy ``peaks'' is determined by the amount of
mass loss adopted.
\begin{figure}
\begin{center}
\includegraphics[width=1.0\textwidth]{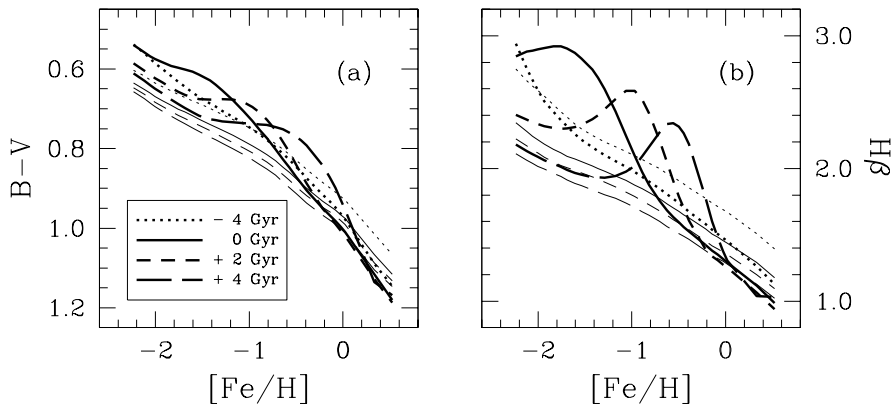}
\end{center}
\caption[]{Two temperature-sensitive 
spectrophotometric indices (a: B-V, b: H$\beta$) as a function
of metallicity and age (individual curves).  The wavy features for
SSPs older than the Galactic GCs are driven by the inclusion of blue
HB stars.}
\end{figure}

\section{Usefulness of Far-UV Photometry}
Perhaps the greatest value that the addition of far-UV photometry
to pre-existing optical data provides is the ability to discriminate 
{\it cleanly} between young ($<$1\,Gyr), intermediate-age
(3$-$5\,Gyr), and old ($>$14\,Gyr), metal-rich 
($-$0.8 $\leq$ [Fe/H] $\leq$ $-$0.4), clusters - this ability
is demonstrated in Figure~2, where it can be seen that a significant
far-UV to optical flux ratio is indicative of either very young or
very old systems (for these metallicities).  In contrast, intermediate-age
clusters are relatively faint in the far-UV.
\begin{figure}
\begin{center}
\includegraphics[width=1.0\textwidth]{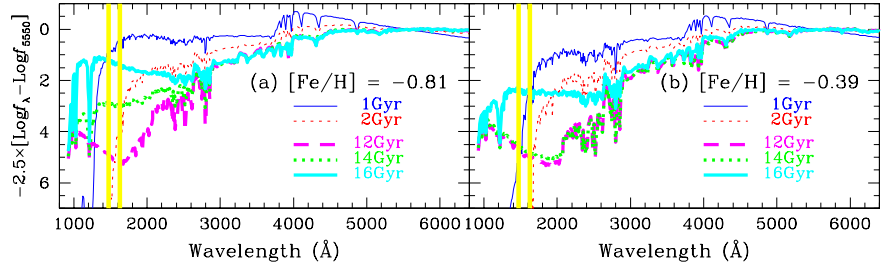}
\end{center}
\caption[]{Selected integrated spectral energy distributions
for metal-rich SSPs.}
\end{figure}

We suggest that the combination of far-UV and optical colours is a most
promising chronometer for SSPs (with the admitted caveat that
there are peculiar GCs such as NGC~6388
and 6441 with populations of extremely blue HB stars - e.g.
Rich et~al. 1997).  {\it If} there are significant age differences
between Milky Way and giant elliptical 
GC systems, however, in the sense of the latter being $>$2~Gyr older (as
suggested by Lee et~al. 2000), then we predict that the majority 
of metal-rich clusters
will be far-UV bright compared to their Galactic counterparts.
In this respect, the outcomes of the recent HST
far-UV photometry of GC systems in Virgo ellipticals
(PIDs\#8643 \& 8725) are highly anticipated; future work with Hubble's
Advanced Camera for Surveys will also be a fruitful avenue of 
research.\footnote{A program to which we would be more than happy to 
contribute ...}  Finally, the upcoming {\tt GALEX} mission will shed
important light on the UV properties of the M31 globular cluster system,
perhaps resolving the controversy regarding their supposed young,
metal-rich, nature (e.g. Burstein et~al. 1984; Barmby \& Huchra 2000).

\section{Nonlinear Colour-Metallicity Relations}
Those Galactic GCs with minimal foreground extinction
(filled circles: inner halo clusters; open circles:
outer halo clusters) are contrasted 
with our models in the ($B$ $-$ $V$)$_{o}$, ($V$ $-$ $I$)$_{o}$,
($M$ $-$ $T_{1}$)$_{o}$, and ($C$ $-$ $T_{1}$)$_{o}$ vs. [Fe/H] planes
in Figure~3.
The data are taken from Harris (1996, 1999 June version) and 
Harris \& Canterna (1977).  We believe this to be one of the first attempts
at employing integrated broadband colours to derive relative age differences
between subpopulations of Galactic GCs.
\begin{figure}
\begin{center}
\includegraphics[width=1.0\textwidth]{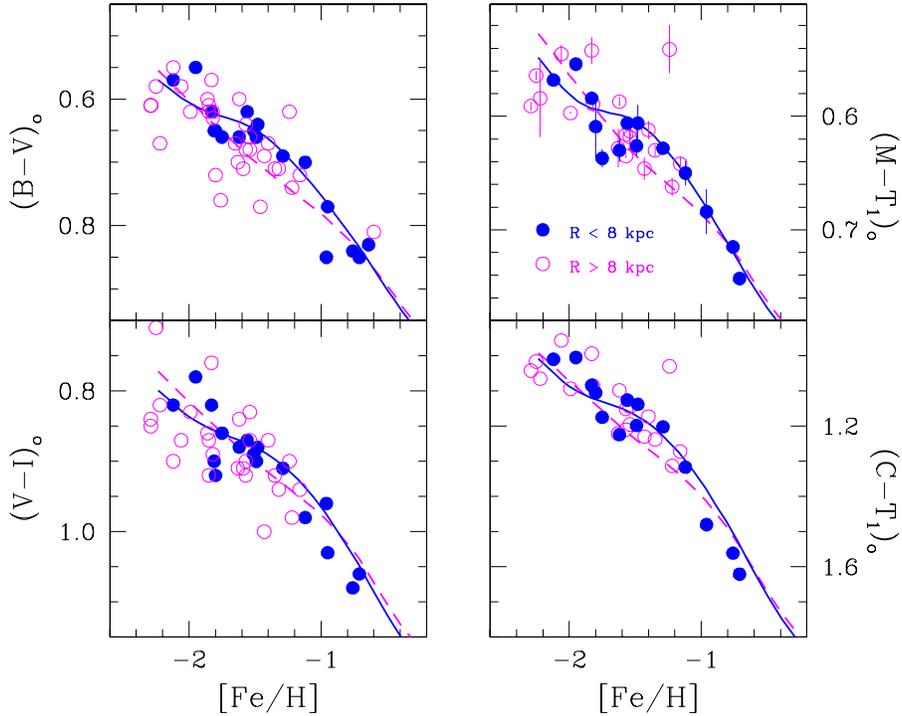}
\end{center}
\caption[]{The relatively low-reddened Galactic GCs
[$E$($B$ $-$ $V$) $<$ 0.2] are used to calibrate our models
in the ($B$ $-$ $V$)$_{o}$, ($V$ $-$ $I$)$_{o}$,
($M$ $-$ $T_{1}$)$_{o}$,
and ($C$ $-$ $T_{1}$)$_{o}$ vs. [Fe/H] planes. The dashed and 
solid lines represent ages of 10~Gyr and 12~Gyr, respectively.}
\end{figure}
It appears that
our models reproduce the differences between inner and outer halo clusters
in the sense that
the inner halo clusters are not only more tightly grouped along
the isochrone than the more scattered outer halo counterparts, 
but also relatively older.
This relative age difference
between inner and outer halo Milky Way GCs is in agreement with that
obtained by Salaris \& Weiss (2002) from their 
homogeneous age dating of 55 clusters (as shown in Figure~4).
The HB morphology is responsible for driving the non-linearity
seen near [Fe/H] $\sim$ $-$1.6, demonstrating that the inclusion of
morphological effects are important if one is to attempt subtle
relative age-dating between SSPs.  
\begin{figure}
\begin{center}
\includegraphics[width=1.0\textwidth]{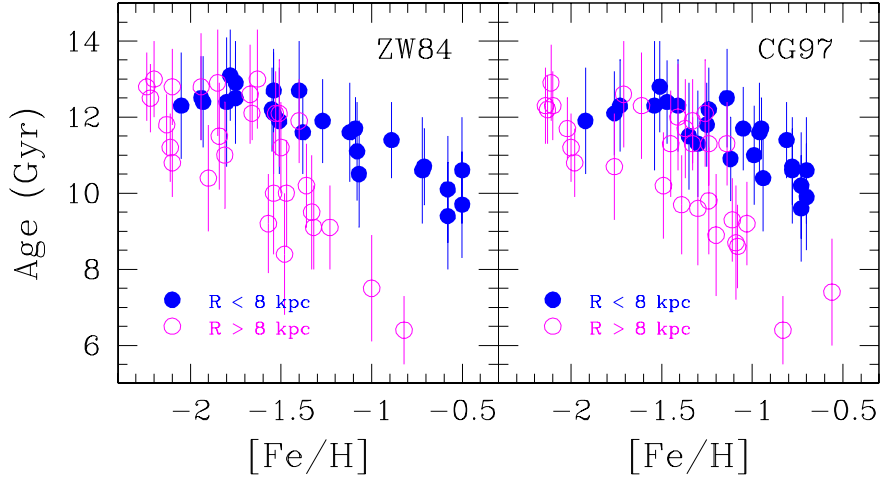}
\end{center}
\caption[]{The relative age differences between inner and outer halo
Milky Way GCs from Salaris \& Weiss (2002).}
\end{figure}

As most of the linear colour-metallicity transformation relations (CMRs)
in the literature are based upon the limited range in colour of Galactic GCs,
caution must be employed when extrapolating their use to redder extragalactic
systems (or if there are sizable age differences amongst globular clusters 
within and between galaxies).
Recently, Kissler-Patig et~al. (1998) provided a set of CMRs using 
the globulars of NGC~1399, but the apparent systematic offset has 
been questioned by Lee et~al. (2002).
It remains to be seen if the bimodal colour distributions 
of extragalactic GC systems can be interpreted
as simple metallicity or age differences, or whether a more complicated
interplay between metallicity and age must be invoked.

\section*{Acknowledgements}

H.-c. Lee sincerely thanks the organizers 
of this beautiful workshop for their hospitality.
Support for this work was provided by the Postdoctoral Fellowship
Program of the Korea Science \& Engineering Foundation, the Creative 
Research Initiatives Program of the Korean Ministry of Science and 
Technology, and the Australian Research Council through its
Linkage International (LX0346898) and Large Grant (A00105171) Programs.

%

\end{document}